\documentclass[aps,pra,twocolumn,showpacs,floatfix,superscriptaddress,10pt]{revtex4-2}

\usepackage[colorlinks,
bookmarksopen,
bookmarksnumbered,
citecolor=teal,
linkcolor=teal,
urlcolor=teal]{hyperref}

\usepackage{xcolor}

\usepackage{amssymb,amsmath,amstext}                
\usepackage{graphicx}                                               
\usepackage{epstopdf}                                               
\usepackage{color}                                                     
\usepackage{bm}                                                        
\usepackage{appendix}                                              
\usepackage[utf8]{inputenc}
\usepackage{latexsym}
\usepackage{braket}
\usepackage{ulem}
\usepackage{siunitx}
\usepackage{psfrag}
\usepackage{mathtools}
\usepackage[english]{babel}
\usepackage[T1]{fontenc}
\usepackage{lmodern}
\usepackage{graphicx}    
\usepackage{orcidlink}

\begin{document}

	\title{Mean-field phase diagrams of spinor bosons in an optical cavity}
	
	\author{Maksym Prodius\orcidlink{0009-0009-1858-882X}}
	\email{maksym.prodius@uj.edu.pl}
	\affiliation{Szko\l{}a Doktorska Nauk \'Scis\l{}ych i Przyrodniczych, Uniwersytet Jagiello\'nski, ulica Stanis\l{}awa \L{}ojasiewicza 11, PL-30-348 Krak\'ow, Poland}
	\affiliation{Instytut Fizyki Teoretycznej, Wydzia\l{} Fizyki, Astronomii i Informatyki Stosowanej, Uniwersytet Jagiello\'nski, \L{}ojasiewicza 11, PL-30-348 Krak\'ow, Poland} 
	
	\author{Mateusz \L{}\k{a}cki\orcidlink{0000-0002-4027-1919}}
	\affiliation{Instytut Fizyki Teoretycznej, Wydzia\l{} Fizyki, Astronomii i Informatyki Stosowanej, Uniwersytet Jagiello\'nski, \L{}ojasiewicza 11, PL-30-348 Krak\'ow, Poland}

	\author{Jakub Zakrzewski\orcidlink{0000-0003-0998-9460}}
	\email{jakub.zakrzewski@uj.edu.pl}
	\affiliation{Instytut Fizyki Teoretycznej, Wydzia\l{} Fizyki, Astronomii i Informatyki Stosowanej, Uniwersytet Jagiello\'nski, \L{}ojasiewicza 11, PL-30-348 Krak\'ow, Poland} 
	\affiliation{Mark Kac Complex Systems Research Center, Jagiellonian University in Krak\'ow, PL-30-348 Krak\'ow, Poland} 
	
	\date{\today} 
	
	\begin{abstract}
		The plethora of possible ground states of spinor bosons placed in an external lattice and a cavity is revisited. We discuss the simplest configuration when the external lattice nodes coincide with the antinodes of the cavity field. We analyze the problem within the grand-canonical mean-field approach, considering both the homogeneous system and the nonhomogeneous case with a harmonic trapping potential. Due to the spin degree of freedom, in the homogeneous case, we treat the system in a twofold manner: we impose the physically relevant total-magnetization constraint, while also discussing the minimization landscape for the full unconstrained problem. In the latter, by combining analytical arguments with numerical calculations based on the Gutzwiller ansatz, we show that the system exhibits two types of magnetic phases: an antiferromagnetic Mott insulator (AFM) and a ferromagnetic density wave (FDW). In addition, two distinct supersolid phases emerge, characterized by different patterns of spin and density imbalances. In the case of zero total magnetization, the FDW phases are replaced by charge density waves (CDWs) that lack net magnetic ordering.  Finally, we establish the phase diagram of the trapped system, providing direct guidance for future experiments.
	\end{abstract}
	
	\maketitle

	\section{Introduction}
	Ultracold atoms placed in an optical cavity provide a unique medium that has been at the center of interest of the quantum optics community for a number of years. Already, early experiments revealed that due to the infinite range of atom-atom interactions mediated by the cavity photons, atoms may self-organize \cite{Domokos02,Black03,Asboth05}. Placing a Bose-Einstein condensate inside an externally pumped cavity enabled condensed matter-type studies by looking at the effective many-body Hamiltonian of the system \cite{Maschler05, Maschler08} in the {\em bad cavity} limit in which the cavity field rapidly adapts to the atomic distribution and may be effectively eliminated. The problem has been analysed in quite a detail using, typically the mean field approach \cite{Larson08, Fernandez10, Dogra16, Niederle16, Chen16} (but also more accurate dynamical bosonic mean field approach \cite{Li13, Panas17}) with increasing interest due to successful experiments \cite{Klinder15, Landig16, Hruby18}. Cavity-induced long-range interactions result in the appearance of charge density wave (checkerboard) insulator or supersolid phases in the superfluid regime, favoring high cavity field intensity. The attempts to faithfully reproduce the experimentally obtained phase diagram for a two-dimensional arrangement resulted in a Quantum Monte Carlo study \cite{Flottat17}, which, to some extent, quite unexpectedly, was largely reproduced by the standard mean-field approach \cite{Himbert19}.
	
	\begin{figure}[!h]
		\centering
		\includegraphics[width=1.0\linewidth]{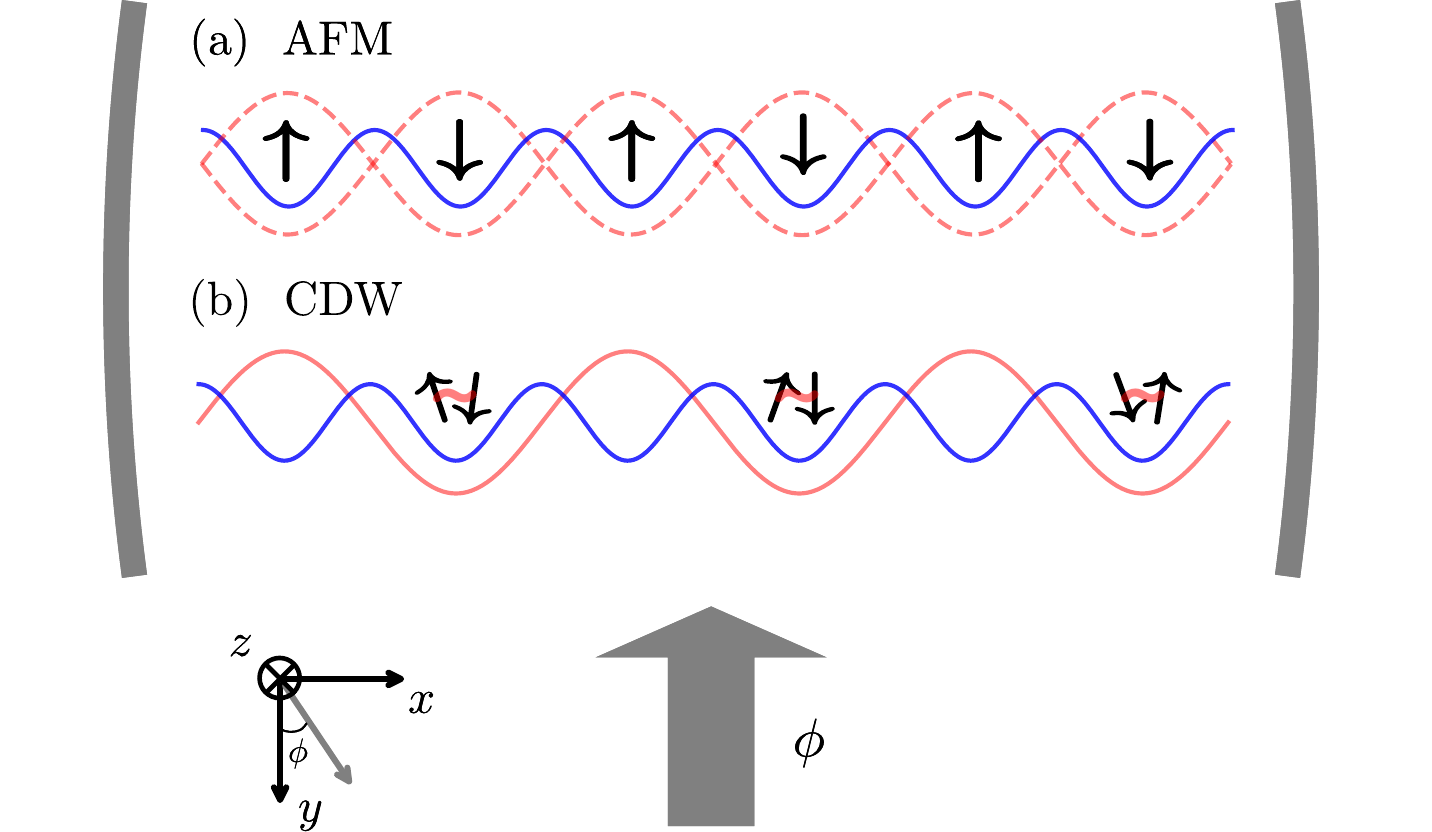}
		\caption{Different phases in a cavity QED setup with spinor bosons. The two-component atoms (represented by arrows) are confined in a two-dimensional optical lattice of size $K = L \times L$ (blue wave) in the $(x,z)$ plane. They interact dispersively with an optical cavity mode (red wave) formed by two mirrors in the $(z,y)$ plane. It is assumed that the lattice nodes coincide with the antinodes of the cavity field. The system is transversely pumped by a laser propagating along the $z$ direction, with polarization angle $\phi$ in the $(x,y)$ plane. By tuning $\phi$, one controls the relative coupling to the vectorial (imaginary) and scalar (real) components of the cavity field. 
			This leads to the emergence of ground-state phases such as the antiferromagnetic Mott insulator (AFM) and the charge density wave (CDW), illustrated for unit filling and zero magnetization as cuts of the two-dimensional lattice in the $(x,y)$ plane in panels (a) and (b), respectively (for the properties of the corresponding order parameters see Table~\ref{tab:phases}). }
		\label{fig:cartoon}
	\end{figure}

	These seminal results were obtained for the situation when the cavity mode and the external pump mode had the same frequency, and the nodes of the cavity field coincided with the minima of the optical lattice potential created by the external pump.  If the frequencies of the cavity and the optical lattice do not match, one may create an effective quasi-periodic potential for atoms, resulting in the appearance of the compressible insulator - a Bose-glass phase \cite{Habibian13, Habibian13b}. Another possible experimental arrangement is when the optical lattice and cavity modes are shifted with respect to each other \cite{Caballero16}, which could lead to the creation of novel phases. That is because the cavity backaction on atoms creates the effective tunnelings between distant sites. As a result, in a one-dimensional lattice, where quasi-exact tensor network-based techniques allow for accurate predictions, one expects to realize a novel self-organized topological insulator \cite{Chanda21} as well as gapless bond-ordered phases \cite{Chanda22}.
	
	Let us repeat, however, that all these predictions are based on the effective Hamiltonian approach resulting from the elimination of the cavity field in the {\em bad cavity} limit. Such a procedure is no longer valid for  {\em good cavities} and too close to the resonance between atoms and the cavity as discussed recently in \cite{Halati20a, Halati20b}. 
	Taking fully into account the dynamics of the atom-cavity model is, however, quite challenging and has been realized using tensor networks in one dimension only. Foa r two-dimensional model, a recent analysis used the approximate truncated Wigner semiclassical approach \cite{Orso25}, revealing long-time metastability in the system. That puts in question tensor network predictions made for a relatively short time evolution \cite{Halati20b}.
	
	Another major modification of the model is extending it beyond the simple two-level atom case. The first experimental attempts \cite{Landini18} (see also \cite{Kroeze18}) already observed the formation of spin textures for multilevel condensates in the cavity, following earlier theoretical studies \cite{Zhou09, Safaei_2013, Padhi14}. This, in turn, stimulated further theoretical work on magnetism in optical cavities (see, e.g., \cite{Mivehvar19, Lozano22}). In particular, \cite{Carl23} considered the extended Bose-Hubbard model \cite{Chanda25} with cavity-mediated interactions for two-component bosons, presenting various possible phases within the mean-field approach. Our work aims to provide a more complete picture of this model within the grand-canonical description and to study the phases that emerge in the presence of an experimentally relevant harmonic trapping potential, offering a comprehensive characterization of the system’s properties within this mean-field framework.
	
	The paper is organized as follows. We define the model in Section~\ref{sec:model}. Then we consider the so-called atomic limit (Section~\ref{sec:atom}) in which we present the possible phases in the limit of vanishing tunnelings. Such an approach allows for an immediate comparison with the spinless case \cite{Himbert19, Dogra16, Sundar16} and reveals the main differences between the standard and spinor cases. The picture obtained suggests a proper selection of parameters of the problem for the subsequent study of mean field phases in the grand canonical ensemble approach presented in Section~\ref{sec:mean}. Typically in the experiments, the relative abundance of different spin components is fixed. Therefore, while usually looking for the unconstrained minimum of the energy, we also consider the case of fixed total magnetization. Finally, in Section~\ref{sec:harm} we consider the experimentally relevant case of harmonic trapping that requires the application of the mean field technique for an inhomogeneous system. As noted already for the scalar case \cite{Sundar16}, the popular local density approximation does not work for infinite range interactions, so such an inhomogeneous Gutzwiller framework is the simplest (although costly) possible approach in the presence of the trap. We present the conclusions in Section~\ref{sec:con}. 
	
	\section{The model}
	\label{sec:model}

	We consider the extended Bose-Hubbard model describing two-component bosons in the optical lattice potential placed in a high-finesse cavity, as depicted in Fig.~\ref{fig:cartoon}. We focus on the two-dimensional geometry of the model originally derived in \cite{Carl23}. The atomic cloud ({envisioned to be formed by two spin components - hyperfine states of $^{87}$Rb atoms}) is pumped from a side by transverse laser light with a tunable polarization angle $\phi$, leading to the emergence of two distinct cavity-mediated interaction terms. The effective Hamiltonian obtained after an adiabatic elimination of the cavity field reads \cite{Carl23}:

	\begin{figure*}[ht!]
		\includegraphics[width=0.9\linewidth]{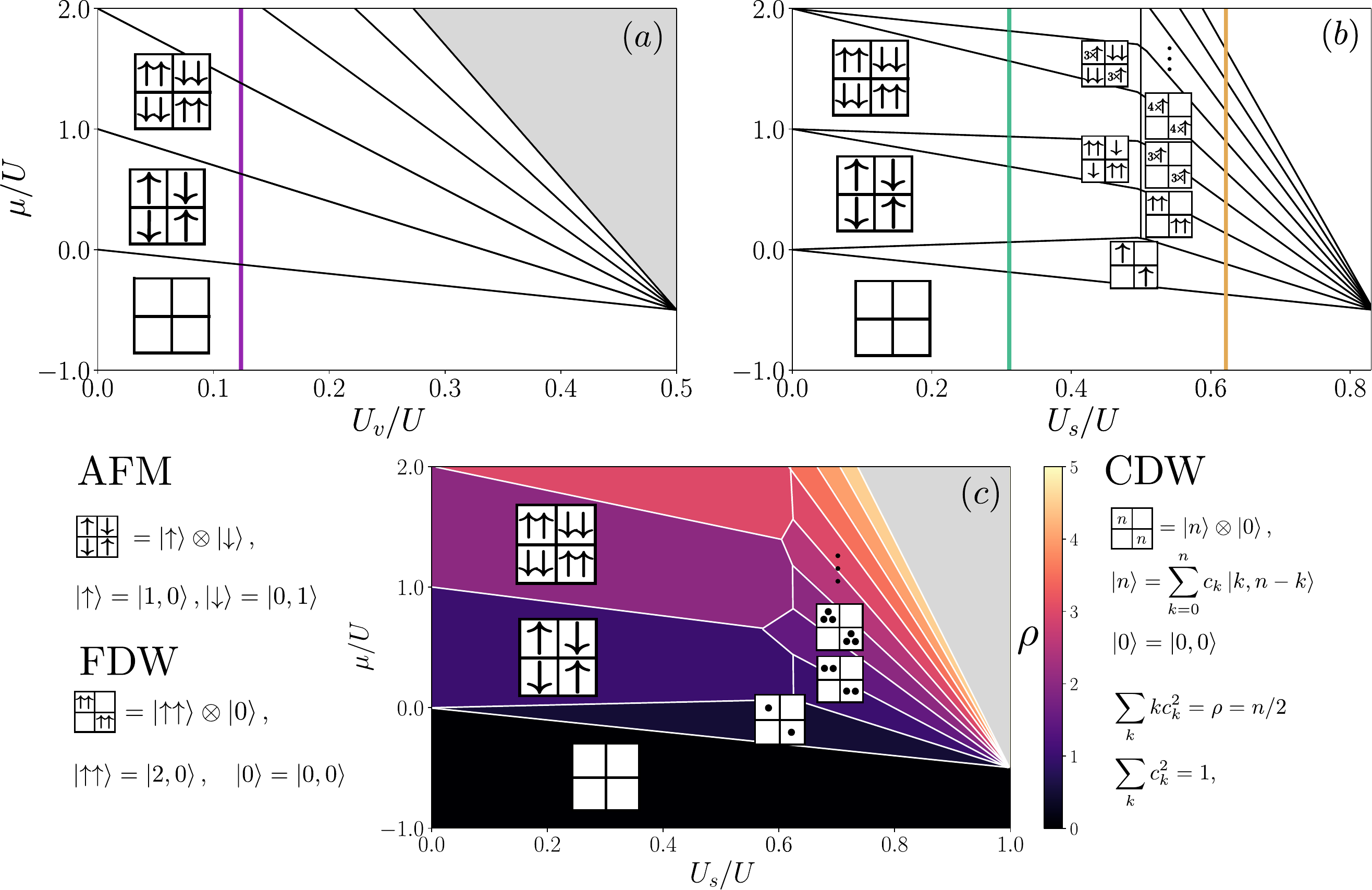}
		
		\caption{Phase diagram of the model (\ref{eq:energy_func}) in the atomic (vanishing tunnelings) limit. The top row depicts the unconstrained regime with $P =0$. Panel (a) illustrates the phases for $U_s / U_v = \frac{1}{5}$, where the transition lines are given by (\ref{eq:mu_Uv}), while panel (b) displays $ U_s / U_v  = 5$ with transition lines defined by (\ref{eq:mu_Us_1}) and (\ref{eq:mu_Us_2}). The subplot (c) presents the case of $ U_s / U_v  = 5$ with a zero-magnetization constraint enforced by $P = 200 U$ and $M_{\textrm{tot}} = 0$. The colormap shows density $\rho$ (\ref{eq:density}) obtained using the Gutzwiller ansatz (\ref{eq:uniform_gutz}). The white solid lines emphasize the transitions between phases. In all the subplots, the gray areas denote the higher-density phases that we do not consider. The colored vertical lines indicate the values of the cavity interaction strength, further examined in Fig.~\ref{fig:mf_lobes} for nonzero tunneling: the purple line corresponds to $U_L = 0.6 U, U_s / U_v = \frac{1}{5}$ (see \eqref{UL}), the green line to $U_L = 0.6 U, U_s / U_v  = 5$, and the orange line to $U_L = 1.2 U, U_s / U_v  = 5$. Square boxes represent the corresponding ground states, with exemplary cases shown in the margins of the figure. }
		\label{fig:atomic_limit}
	\end{figure*}

	\begin{equation}
		\begin{aligned}
			\hat{H}
			&= - t  \sum_{\left<i, j \right>}^{K} \sum_{\sigma \in\{\uparrow, \downarrow\}} \left(\hat{b}_{i, \sigma}^{\dagger}\hat{b}_{j, \sigma} + h.c \right)  \\
			&\quad
			+ \frac{U}{2} \sum_{i=1}^{K} \sum_{\sigma  \in\{\uparrow, \downarrow\}} \hat{n}_{i, \sigma} (\hat{n}_{i, \sigma} - 1) + U_{12} \sum_{i=1}^{K} \hat{n}_{i, \uparrow} \hat{n}_{i, \downarrow} \\ 
			&\quad
			- \frac{U_s}{K}\hat{\Theta}_s^2 - \frac{U_{v}}{K}\hat{\Theta}_v^2\,.
			\label{eq:ham}
		\end{aligned}
	\end{equation}
	The operators $\hat{b}^{\dagger}_{i, \sigma} (\hat{b}_{i, \sigma})$ create (annihilate) bosons with spin $\sigma \in \{\uparrow, \downarrow\}$ at site $i = (i_x, i_z)$ {of the lattice with $K = L \times L$ lattice sites}, while $\hat{n}_{i, \sigma} = \hat{b}^{\dagger}_{i, \sigma} \hat{b}_{i, \sigma}$ are the corresponding particle number operators. { $\sum_{i=1}^K$ stands for the sum over all sites $\sum_{i_x=1}^L\sum_{i_z=1}^L$}. The first term in \eqref{eq:ham} describes the nearest-neighbor hopping with amplitude $t>0$. Repulsive onsite contact interactions are proportional to $U > 0$ and $U_{12} > 0$. Below, we focus on the experimentally relevant case $U_{12} = U$. Finally, the last row of \eqref{eq:ham} corresponds to the cavity-mediated,  scalar and vectorial interaction terms:
	\begin{equation}
		\begin{aligned}
			\hat{\Theta}_s =  \sum_{i=1}^{K} (-1)^{|i|} (\hat{n}_{i, \uparrow} + \hat{n}_{i, \downarrow} ), \\
			\hat{\Theta}_v = \sum_{i=1}^{K} (-1)^{|i|} (\hat{n}_{i, \uparrow} - \hat{n}_{i, \downarrow} )\, ,
		\end{aligned}
	\end{equation}
	where $|i| =i_x + i_z$. The scalar term, proportional to $U_s > 0$ drives the site occupation imbalance, acting like its single-species counterpart \cite{Landig16}. In contrast, the vectorial term ($U_v > 0$) tends to induce spin imbalance instead. The relative role of these two terms  is tunable experimentally via the change of the cavity polarization angle $\phi$ \cite{Carl23}:
	\begin{equation}
		U_s = U_L \mathrm{cos}^2 \phi, \quad  U_v = U_L \xi^2 \mathrm{sin}^2 \phi, \label{UL}
	\end{equation}
	where $\xi = \frac{\alpha_v}{2 \alpha_s}$ is a constant proportional to the ratio of vectorial and scalar polarizabilities of the underlying atoms. In this work we assume $\xi = 0.464$, corresponding to $^{87}$Rb \cite{Carl23}, while the overall strength $U_L$ can be changed by, e.g., adjusting the detuning of the pumping laser.

	Within the mean-field analysis, we focus on the ground-state properties of the grand-canonical counterpart to the Hamiltonian \eqref{eq:ham}:
	\begin{equation}
		\hat{H}_{GC} = \hat{H} -  \mu \sum_{i=1}^{K}\sum_{\sigma \in \{ \uparrow, \downarrow\}} \hat{n}_{i, \sigma},
		\label{eq:grand}
	\end{equation}
	where we assume the same chemical potential $\mu$ for both spin components. We look for the ground state $ \ket{\Psi}$ by minimizing the energy functional
	\begin{equation}
		\mathcal{E} (\ket{\Psi}) =  \braket{\Psi | \hat{H}_{GC} | \Psi} + \frac{P}{K} \left(\braket{\Psi | \hat M | \Psi} - M_{\textrm{tot}} K \right)^2 \label{eq:energy_func},
	\end{equation}
	where an additional magnetization penalty (soft constraint) term proportional to the positive number $P$ is added, with $ \hat{M} = \sum_{i=1}^K (\hat{n}_{i, \uparrow} -  \hat{n}_{i, \downarrow})$ being the total magnetization operator and $M_{\textrm{tot}}$ set to the desired magnetization per site.
	
	In the following, we consider two cases. Assuming $P=0$, we determine the zero-temperature phase diagrams of the unconstrained grand-canonical description of the problem. Then we compare these ground states to the case of sufficiently large, positive $P$ and $M_{\textrm{tot}} = 0$, which enforces an equal number of spin-up and spin-down bosons (zero total magnetization), and briefly comment on other choices for the net magnetization parameter ($M_{\textrm{tot}} \neq 0$).

	\section{Homogeneous Atomic limit}
	\label{sec:atom}
	
	{To gain insight into the possible insulating phases of the infinite-lattice model, we begin by neglecting hopping terms and consider the so-called atomic limit. In this limit, the general constrained problem (\ref{eq:energy_func}) cannot be solved by simple minimization in the space of Fock states. Therefore, we formulate the problem using a uniform Gutzwiller ansatz with a two-site unit cell defined by odd and even sites, reflecting the structure of the Hamiltonian (\ref{eq:grand}):}
	\begin{equation}
		\left|\Psi \right> = \prod_{i=1}^{K/2} \left(\left|\phi_e \right>_{2i} \otimes \left|\phi_o \right>_{2i - 1} \right) ,
		\label{eq:uniform_gutz}
	\end{equation}
	\begin{equation*}
		\left|\phi_i \right> = \sum_{n=0}^{n_{\textrm{max}}} \sum_{m=0}^{m_{\textrm{max}}} g_i(n, m) \left|n,m\right>_i,
	\end{equation*}
	where $\left|n,m\right>_i$ denote Fock states accounting for the two spin components with $n$ bosons with spin up ($\uparrow$) and $m$ particles with spin down ($\downarrow$).  Here $i = e, o$ and the cut-offs $n_{\textrm{max}} = m_{\textrm{max}} = 10$ are assumed for the local Hilbert spaces. Generally, the Gutzwiller coefficients $g_i(n, m)$ are assumed to be real and are obtained via numerical minimization of the energy density $\varepsilon(g_e, g_o) = 2 \mathcal{E}(g_e, g_o) / K $ (\ref{eq:energy_func}); further details of this procedure are provided in Appendix~\ref{app:hom_gutz}. 
	
	\subsection{Unconstrained grand-canonical phases}
	We start with the $P = 0$ case in (\ref{eq:energy_func}). {Without the constraint, we solve the pure Hamiltonian problem, which is diagonal in the particle number operators for vanishing tunneling. Then, the problem can be solved analytically by minimizing the energy in the space of Fock states of the form $\ket{\phi_e}_{2i} = \ket{n_{e, \uparrow}, n_{e, \downarrow}}$ and $\ket{\phi_o}_{2i-1} = \ket{n_{o, \uparrow}, n_{o, \downarrow}}$, where $n_{i, m}$ are the four corresponding occupation numbers.} 
	
	The cavity-mediated terms decouple naturally and become proportional to $\theta_s^2$ and $\theta_v^2$, where we define
	\begin{equation}
		\begin{aligned}
			\theta_s = \braket{\hat{n}_{e, \uparrow}} + \braket{\hat{n}_{e, \downarrow}} -  \braket{\hat{n}_{o, \uparrow}}
			- \braket{\hat{n}_{o, \downarrow}}, \\
			\theta_v = \braket{\hat{n}_{e, \uparrow}} - \braket{\hat{n}_{e, \downarrow}} -  \braket{\hat{n}_{o, \uparrow}} + \braket{\hat{n}_{o, \downarrow}}. 
			\label{eq:thetas}
		\end{aligned}
	\end{equation}
	Additionally, we denote the total density as
	\begin{equation}
		\rho = \frac{\braket{\hat{n}_{e, \uparrow}} + \braket{\hat{n}_{o, \uparrow}} + \braket{\hat{n}_{e, \downarrow}}   + \braket{\hat{n}_{o, \downarrow}}}{2}.
		\label{eq:density}
	\end{equation}
	The energy density $\varepsilon(\ket{\Psi} ) =  \mathcal{E}(\ket{\Psi} ) / K$ (\ref{eq:energy_func}) may be expressed in the following convenient form (see the similar result for the single spin model in\cite{Himbert19}):
	
	\begin{equation}
		\begin{aligned}
			\varepsilon( \rho, \theta_s, \theta_v) &= \frac{U}{2} \rho (\rho - 1) - \mu \rho  \\
			&\quad
			+ \left(\frac{U}{2} - U_s \right) \frac{\theta_s^2}{4}- U_v \frac{\theta_v^2}{4}.
			\label{eq:atomic_limit}
		\end{aligned}
	\end{equation}
	By identifying the  Fock states that minimize  (\ref{eq:atomic_limit}), we find the ground state properties of the model in the atomic limit, hence determining the possible insulating phases in the grand-canonical description. The detailed derivation of (\ref{eq:atomic_limit}) is given in the Appendix~\ref{app:atomic_limit}.

	We first consider the regime $U_v > U_s$. In this case, the ground-state configuration is a Fock state that maximizes the spin imbalance $\theta_v$, namely $|\theta_v| = 2\rho$, while the density imbalance $\theta_s=0$. 
	
	As a result, we obtain the ground-state phase diagram shown in Fig.~\ref{fig:atomic_limit}(a) in the $\mu / U - U_v / U$ space, defined by the relation
	
	\begin{equation}
		\left(U - 2U_v \right) \rho - U + U_v < \mu < \left(U - 2U_v \right) \rho - U_v.
		\label{eq:mu_Uv}
	\end{equation}
	As can be seen, for $U_v < U / 2$  the only possible phases are commensurate antiferromagnetic Mott insulators (AFMs), in which the particle number increases abruptly by two as the chemical potential is raised, with opposite spins occupying opposite sites. Note that for $U_v > U / 2$ the description by (\ref{eq:atomic_limit}) is no longer valid.
	
	The regime $U_v < U_s$ exhibits a more intricate structure. From (\ref{eq:atomic_limit}), it is apparent that, depending on the value of $U_s$, two distinct situations may arise. 
	
	When $U_s < U / 2$, the system displays the same AFM phases as in the $U_v > U_s$ regime; however, additional intermediate phases appear, characterized by a density imbalance pattern with $|\theta_s| = 1$ and half-integer densities $\rho$ - see Fig.~\ref{fig:atomic_limit}(b). Phases that simultaneously exhibit scalar (density) and vectorial (spin) order are referred to as a ferromagnetic density wave (FDW). 
	
	When $U_s > U / 2$, minimizing (\ref{eq:atomic_limit}) requires both cavity terms to be maximized, $|\theta_s| = |\theta_v| = 2 \rho$. This again leads to FDW phases, but with one site fully occupied and the other empty.
	
	In both cases, the phase boundaries are determined by the chemical-potential relations (\ref{eq:mu_Us_1}) and (\ref{eq:mu_Us_2})
	\begin{multline}
		\left(U - 2 U_v\right) \rho - U + \frac{U_s + U_v}{2}  < \mu \\ <   \left(U - 2 U_v\right)\rho -  \frac{U_s + U_v}{2},
		\label{eq:mu_Us_1}
	\end{multline}
	\begin{multline}
		2  \left(U - 2 (U_s+ U_v) \right)\rho - U + \frac{U_s + U_v}{2}  < \mu \\   < 2 \left(U - 2 (U_s + U_v)\right)\rho - \frac{U_s + U_v}{2},
		\label{eq:mu_Us_2}
	\end{multline}
	and the corresponding phase diagram is shown in Fig.~\ref{fig:atomic_limit}(b). It is worth noting that the point at which this analysis becomes invalid is $U_s = U$ when $U_v = 0$. For nonzero values of $U_v$, this point shifts to the left, while in the limit $U_v$ approaching $U_s$, everything shrinks to the same phase diagram as for $U_v > U_s$. This behavior is governed by the condition that, for Eq.~(\ref{eq:atomic_limit}) to remain valid in this regime, the cavity-mediated strengths must satisfy $U_s + U_v < U$.

	\begin{figure*}
		\centering
		\includegraphics[width=1.0\linewidth]{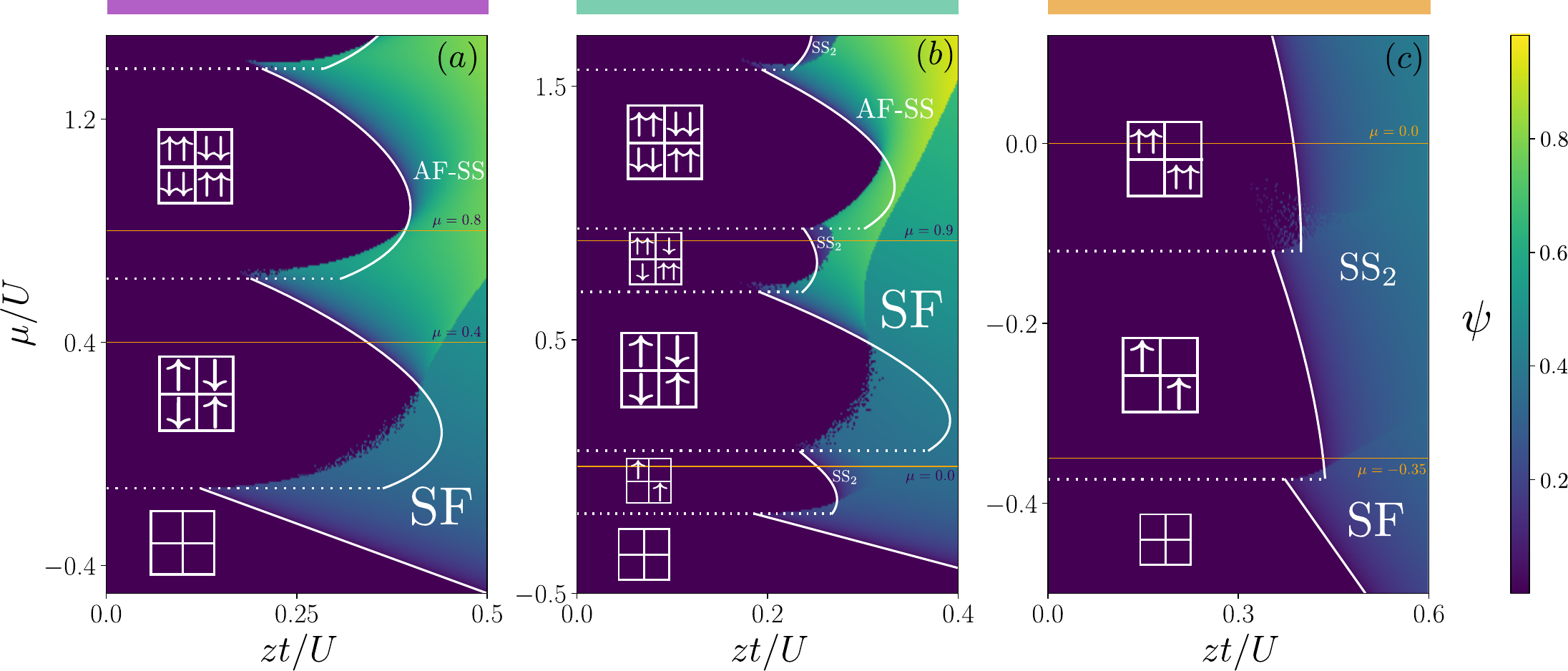}
		
		\caption{Mean-field phases of the unconstrained homogeneous system in the ($z t / U$, $\mu / U$) plane. The underlying color map represents the $\psi$ order parameter obtained numerically within the Gutzwiller approximation, while white solid lines correspond to the phase boundaries calculated perturbatively (dashed lines represent the continuation of the atomic limit solutions). (a)  $U_s / U_v =\frac{1}{5}$ and $U_L = 0.6 U$,  (b)  $U_s / U_v = 5$ and $U_L = 0.6 U$ and (c)  $U_s / U_v = 5$ and $U_L = 1.2 U$. Yellow horizontal lines correspond to the values of the chemical potential analyzed in Fig.~\ref{fig:diagram_cuts}. 
			The square boxes represent the underlying AFM and FDW ground states. The properties of all phases are summarized in Table~\ref{tab:phases}.}
		\label{fig:mf_lobes}
		
	\end{figure*}

	In all of the above cases for $U_L \rightarrow 0$ ($U_s =0,  U_v = 0$) we have a standard Bose-Hubbard Mott Insulator relation $\frac{\mu}{U} \in \left[ \rho - 1, \rho\right]$ for integer $\rho$. 
	
	{
		\subsection{Phases with {fixed total magnetization}}
		
		{We consider now the case of fixed total magnetization, formally realized by setting $P > 0$ in (\ref{eq:energy_func}). We focus on a typical example of equally populated up and down spin components resulting in a zero-magnetization regime, $M_{\textrm{tot}} = 0$.}
		For the dominant vectorial interactions, the zero-magnetization penalty term is trivially satisfied, so the phase diagram is equivalent to Fig.~\ref{fig:atomic_limit}(a). Differences become apparent only in the opposite regime, $U_s > U_v$. We present the zero-magnetization Gutzwiller results in Fig.~\ref{fig:atomic_limit}(c), where we plot the density (\ref{eq:density}) of the resulting ground states.
		
		First, we observe that the phase boundaries shift to the right compared to the unconstrained analytical predictions, although the overall phase diagram structure remains similar. Another prominent difference is that AFM phases of different densities are now directly connected, without any intervening intermediate phases. Second, some of the underlying ground states also change. As expected, the AFM phases remain unchanged, whereas the FDW phases, which violate the constraint, are replaced by different ground states. Specifically, from the Gutzwiller wavefunctions, we can read out the explicit forms of these new states: $\ket{\phi}_{e} = \ket{0, 0}$ and $\ket{\phi}_{o} = \sum_{k}^{2\rho} c_k \ket{k, 2\rho - k}$ or vice versa. The Gutzwiller coefficients are constrained by $\sum_{k}^{2\rho} c_k^2 = 1$ and $\sum_{k}^{2\rho} k c_k^2 = \rho$ for $\rho \in \{0.5, 1, 1.5,... \}$. These solutions are characterized by $|\theta_v| = 0$ and $|\theta_s| = 2 \rho$, representing, on a mean-field level, checkerboard density wave states, where one sublattice is always empty and the other contains superpositions of spin components that satisfy the zero-magnetization condition. These superpositions are degenerate in energy, reflecting the underlying symmetry for $U_{12} = U$. Note that there are even more degenerate solutions beyond the two-site ansatz. For this reason, we refer to these ground states as charge density waves (CDWs), reflecting no underlying magnetic pattern on the occupied sublattice.

	}

	\section{Homogeneous Mean field phases}
	\label{sec:mean}
	Next, we turn on the tunneling terms. To capture the full qualitative behavior of the model for $U = U_{12}$, we consider three representative cases of cavity-mediated interactions highlighted by colored lines in Fig.~\ref{fig:atomic_limit}. We approach this problem from two perspectives. In subsection~\ref{subsect:perturb}, we build on the atomic-limit solutions and determine the insulating phases using perturbation theory {in the unconstrained grand-canonical case}. In subsection~\ref{subsect:gutzwiller}, we compare these results with nonperturbative calculations based on the numerical Gutzwiller ansatz. {Finally, in subsection~\ref{subsect:zero_magnetization}, we investigate what will change if we include the constant magnetization penalty.} 
	
	The different ground-state phases are distinguished by the imbalance order parameters $\theta_s$ and $\theta_v$ (see ~(\ref{eq:thetas})) and by the superfluid order parameter $\psi = \sum_{i,\sigma} \left| \braket{\Psi| \hat{b}_{i,\sigma} |\Psi } \right| / 4$. From this point on, we classify phase transitions as continuous or discontinuous depending on whether the relevant order parameters vary continuously or discontinuously across the transition.

	\begin{figure}[ht]
		\centering
		\includegraphics[width=1.0\linewidth]{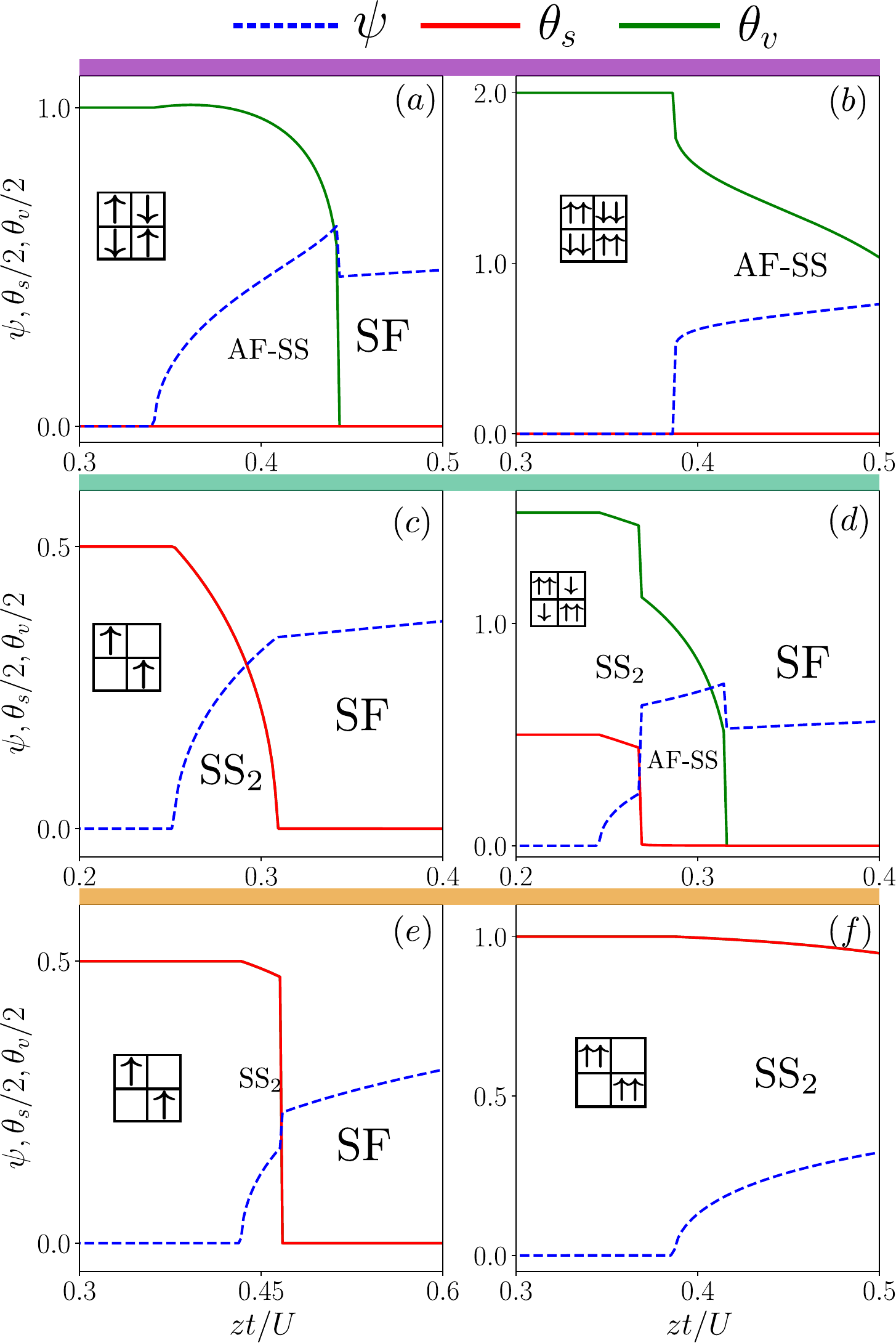}
		\caption{Horizontal cuts of the phase diagrams in Fig.~\ref{fig:mf_lobes} for fixed chemical potential. The first row corresponds to the cavity interaction parameters as in Fig.~\ref{fig:mf_lobes}(a) with $\mu  / U = 0.4 $ in subplot (a) and $\mu  / U= 0.8 $ in subplot (b). The second row corresponds to Fig.~\ref{fig:mf_lobes}(b) with $\mu / U = 0.0 $ in subplot (c) and $\mu / U = 0.9 $ in subplot (d). The last row corresponds to Fig.~\ref{fig:mf_lobes}(c) with $\mu / U = -0.35$ in subplot (e) and $\mu  / U= 0.0$ in subplot (f).}
		\label{fig:diagram_cuts}
	\end{figure}

	\subsection{Perturbative predictions}
	\label{subsect:perturb}
	
	Taking the no-tunneling limit as the unperturbed Hamiltonian, we determine the phase boundaries using perturbation theory~\cite{Oosten01}, treating the mean-field–decoupled hopping terms as perturbations. Within second-order perturbation theory, and assuming continuous transitions, the boundaries of the insulating phases are given by
	\begin{multline}
		\frac{1}{t_\sigma^2} = z^2 \left(\frac{n_{e,\sigma} + 1}{\Delta E^{(0)}_{n_{e, \sigma}+1}} + \frac{n_{e,\sigma}}{\Delta E^{(0)}_{n_{e,\sigma}-1}}\right)\\
		\times \left(\frac{n_{o,\sigma} + 1}{\Delta E^{(0)}_{n_{o, \sigma}+1}} + \frac{n_{o,\sigma}}{\Delta E^{(0)}_{n_{o, \sigma}-1}}\right),
		\label{eq:perturb}
	\end{multline}
	where $z = 4$ is the coordination number, $\sigma \in \{\uparrow, \downarrow\}$, and $\Delta E^{(0)}_{n_{i, \sigma} \pm 1}$ denotes the particle/hole excitation energies of the unperturbed Hamiltonian. We obtain two solutions, $t_{\uparrow}(\mu)$ and $t_{\downarrow}(\mu)$. { Given an appropriate choice of unit cells, symmetry ensures identical transitions for both. } In Fig.~\ref{fig:mf_lobes}, these results are shown as white solid lines. {For nonzero tunneling, the AFM and FDW phases obtained in the atomic limit evolve into characteristic insulating lobes in the $(zt/U, \mu/U)$ plane. In all considered cases (a), (b), and (c), at small tunneling, the solutions for lobes with different densities overlap and, more importantly, do not match the atomic-limit predictions at $t = 0$, signaling first-order transitions between these insulating phases. Therefore, in the vicinity of zero tunneling, we instead plot the continuation of the atomic-limit solutions as dashed lines. The transitions from the insulating phases to the non-insulating regime, denoted by the solid lines, are continuous by construction.}

	\subsection{Nonperturbative results}
	\label{subsect:gutzwiller}

	\begin{table}
		\centering
		\renewcommand{\arraystretch}{1.2}
		\begin{tabular}{|l|l||l|l|l|}
			\hline
			Phases & Acronyms & $\psi$ & $|\theta_s|$ & $|\theta_v|$ \\
			\hline\hline
			Antiferromagnetic Mott insulator  & AFM & $0$ & $0$ & $2 \rho$ \\
			Ferromagnetic Density Wave & FDW & $0$ & $2 \rho$ & $2 \rho$ \\
			AF Lattice Supersolid & AF-SS & $\neq 0$ & $0$ & $\neq 0$ \\
			Type-2 Lattice Supersolid & SS$_2$ & $\neq 0$ & $\neq 0$ & $\neq 0$ \\
			Superfluid & SF & $\neq 0$ & $0$ & $0$\\
			\hline 
			Charge Density Wave & CDW & $0$ & $2 \rho$ & $0$ \\
			Lattice Supersolid & SS & $\neq 0$ & $\neq 0$ & $0$ \\
			\hline
			
		\end{tabular}
		\caption{The mean-field ground state phases for the homogeneous system, their acronyms, and the properties of the corresponding order parameters. $\rho$ is a density of particles defined by (\ref{eq:density}).}
		\label{tab:phases}
	\end{table}

	We use the Gutzwiller ansatz introduced in the previous section (\ref{eq:uniform_gutz}). Additionally, to properly determine the grand canonical ground states in the FDW phase, we include a small symmetry-breaking field in the minimization functional. Specifically, we introduce a weak magnetic field $h = 10^{-4}U$ through the term $h \sum_i (\hat{n}_{i, \uparrow} - \hat{n}_{i, \downarrow})$, which lifts the degeneracy and ensures the correct evaluation of the order parameter $\theta_v$.

	To facilitate comparison with the perturbative predictions, we present the numerically obtained superfluid order parameter $\psi$ in Fig.~\ref{fig:mf_lobes}. Furthermore, Fig.~\ref{fig:diagram_cuts} displays the behavior of the relevant order parameters along selected horizontal cuts of the phase diagram, providing a more detailed view of the phases and clarifying the nature of the transitions. For convenience, the characteristics of the phases discussed in this section are summarized in Tab.~\ref{tab:phases}.

	We now turn to the case of dominant vectorial interactions shown in Fig.~\ref{fig:mf_lobes}(a). In addition to the AFM phases, we observe a superfluid (SF) phase and an antiferromagnetic supersolid phase, denoted AF-SS. Both phases are characterized by a nonzero  $\psi$, while AF-SS additionally exhibits $\theta_v \neq 0$. The transition between the SF and AF-SS phases is {discontinuous} in $\psi$. In contrast, transitions involving the insulating lobes are more intricate than suggested by the perturbative predictions. In particular, transitions {roughly} from the upper (lower) parts of the lobes appear to be continuous (discontinuous), as illustrated by comparing Fig.~\ref{fig:diagram_cuts}(a) and Fig.~\ref{fig:diagram_cuts}(b). It is also worth noting that the small-tunneling behavior of the Gutzwiller solutions is in good agreement with the dashed white lines. However, the overall placement of the lobes, especially for the first lobe, differs from the perturbative results. The nonperturbative calculation predicts these transitions to occur at smaller values of $z t / U$ and to be discontinuous. We note that for lobes with higher densities (not shown), this discrepancy is much smaller.
	
	Figure~\ref{fig:mf_lobes}(b) shows the regime where scalar interactions are dominant, but $U_s < U / 2$. The overall phase diagram is similar to that in panel (a); however, as predicted from the atomic-limit analysis, new insulating phases appear between the AFM lobes, characterized by both $\theta_s \neq 0$ and $\theta_v \neq 0$. {These FDW lobes are surrounded by an additional supersolid phase, denoted as SS$_2$. It is characterized by $\theta_s \neq 0$, $\theta_v \neq 0$, and $\psi \neq 0$. The continuous transition from the FDW with $\rho= 1/2$ to SS$_2$ and then from SS$_2$ to SF is shown in Fig.~\ref{fig:diagram_cuts}(c). For higher values of $\mu$ the FDW with higher half-integer densities, e.g $\rho = 3/2$ seen in the Fig.~\ref{fig:mf_lobes}(b), are also possible. With increased tunneling, the system undergoes a continuous transition into the SS$_2$ phase. The transition from SS$_2$ to AF-SS is discontinuous, signaled by a jump of the superfluid order parameter and both $\theta$'s, as shown in Fig.~\ref{fig:diagram_cuts}(d). Once again, for both the AFM and FDW lobes, the transitions are no longer continuous where they do not coincide with the perturbative white lines.}
	
	Finally, Fig.~\ref{fig:mf_lobes}(c) covers the case $U_s > U_v$ and $U_s > U / 2$. In contrast to the previous two configurations, the perturbative phase boundaries are visually in better agreement with the Gutzwiller results. {  In this regime, only FDW insulating lobes are present, surrounded by the SS$_2$ and SF phases. Interestingly, unlike in Fig.~\ref{fig:diagram_cuts}(c), the transition from SS$_2$ to SF here for the chosen $\mu$ appears to be discontinuous (Fig.~\ref{fig:diagram_cuts}(e)). The FDW-SS$_2$ transition remains continuous (Fig.~\ref{fig:diagram_cuts}(e-f)).}

	\subsection{Magnetization constraint}
	\label{subsect:zero_magnetization}
	
	\begin{figure}[t!]
		\centering
		\includegraphics[width=1.0\linewidth]{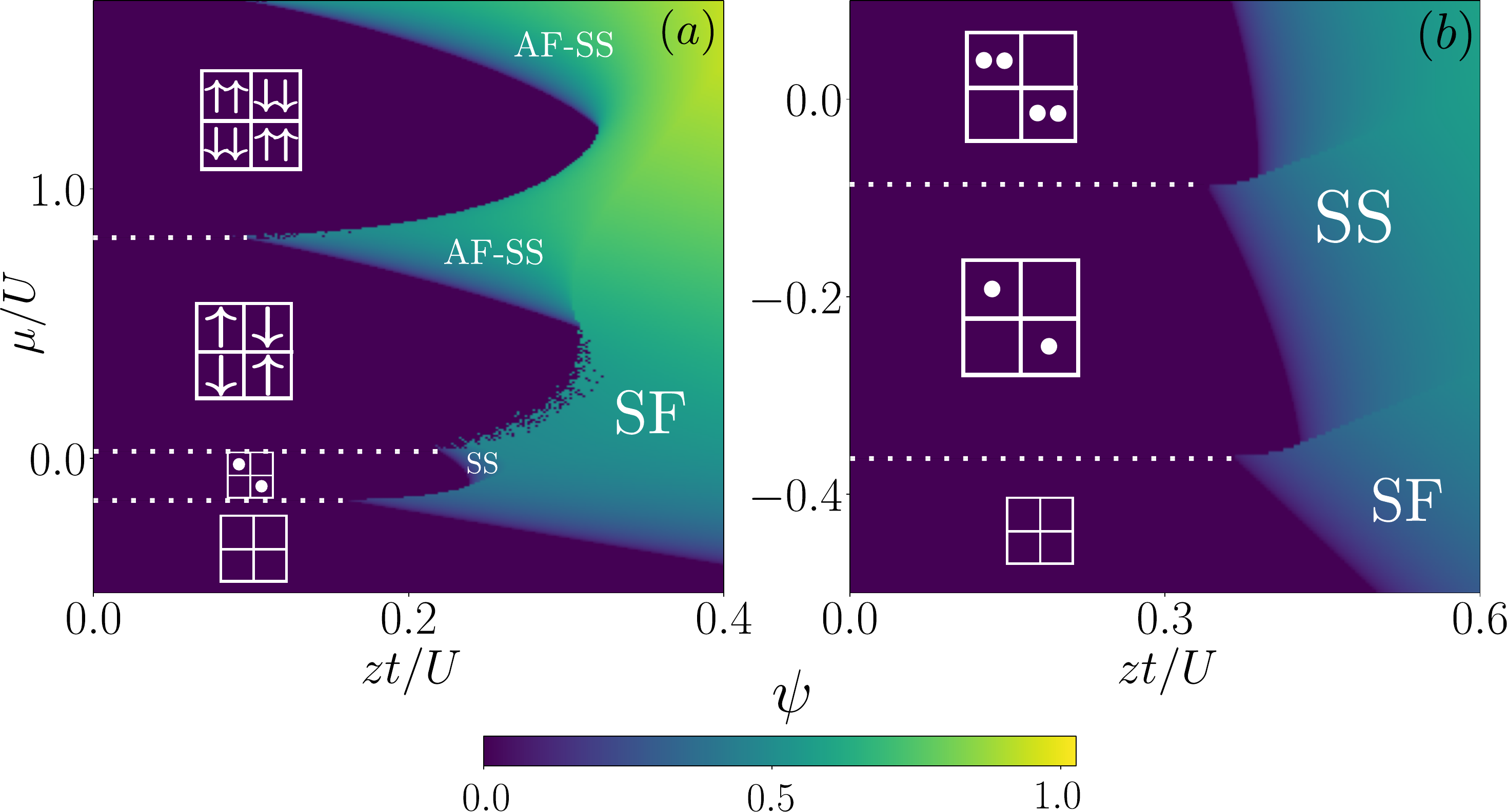}
		\caption{Mean-field phases of the homogeneous system with zero total magnetization ($P = 200U, M_{\textrm{tot}} = 0$) in the ($t / U$, $\mu / U$) plane (compare with the unconstrained diagram in Fig.~\ref{fig:mf_lobes}). Panel (a) shows the case for $U_s / U_v = 5$ and $U_L = 0.6 U$, and panel (b) for $U_s / U_v = 5$ and $U_L = 1.4 U$. {The white dashed lines show the transitions between lobes of different density obtained from calculations in the atomic limit.} } 
		\label{fig:constrained_mean_field}
	\end{figure}
	
	By introducing a strictly positive $P$ into Eq.~(\ref{eq:energy_func}), we restrict the system to phases with the same total magnetization. In Fig.~\ref{fig:constrained_mean_field}, we present the finite tunneling phase diagrams for the zero-magnetization constraint ($M_{\textrm{tot}} = 0$). We focus exclusively on the regime of dominant scalar interactions, as the $U_v > U_s$ regime yields a phase diagram identical to the $P=0$ case shown in Fig.~\ref{fig:mf_lobes}(a).
	
	{Under the enforced zero-magnetization constraint, the ground-state phase diagram for $U_L = 0.6 U, U_s/U_v = 5$ in Fig.~\ref{fig:mf_lobes}(b) transforms into Fig.~\ref{fig:constrained_mean_field}(a). The first notable difference is that the insulating lobe with $\rho=1/2$ now hosts CDW instead of FDW states, as discussed in the atomic limit section. Second, the FDW lobes with half-integer densities greater than $\rho =1 / 2$ disappear completely, as do the SS$_2$ states. The latter either disappear or are replaced by the standard supersolid regime SS already observed for the one-component cavity model \cite{Landig16}, which is characterized by $\theta_v = 0$, $\theta_s = 0$, and $\psi \neq 0$.
		
		Comparing Fig.~\ref{fig:mf_lobes}(c) with Fig.~\ref{fig:constrained_mean_field}(b) for $U_L = 1.4 U, U_s/U_v = 5$ reveals that the FDW lobes are replaced by CDW lobes. 
		As before, the SS phase appears in the place of SS$_2$.
		
		One should also address what happens for choices of total magnetization other than $M_{\textrm{tot}} = 0$. A glimpse of this information is already encoded in the unconstrained phase diagrams (Fig.~\ref{fig:mf_lobes}). Initializing all bosons with spins polarized in a single direction yields, for example, the FDW phases like in Fig.~\ref{fig:mf_lobes}(c) - resembling the behavior of the one-component model. If a magnetization that allows for FDW states with $\rho = 3/2$ is considered, the novel SS$_2$ phase also emerges alongside. A more detailed analysis of other relevant magnetizations may be explored in future work.

		\begin{figure*}[!ht]
			\centering
			\includegraphics[width=1.0\linewidth]{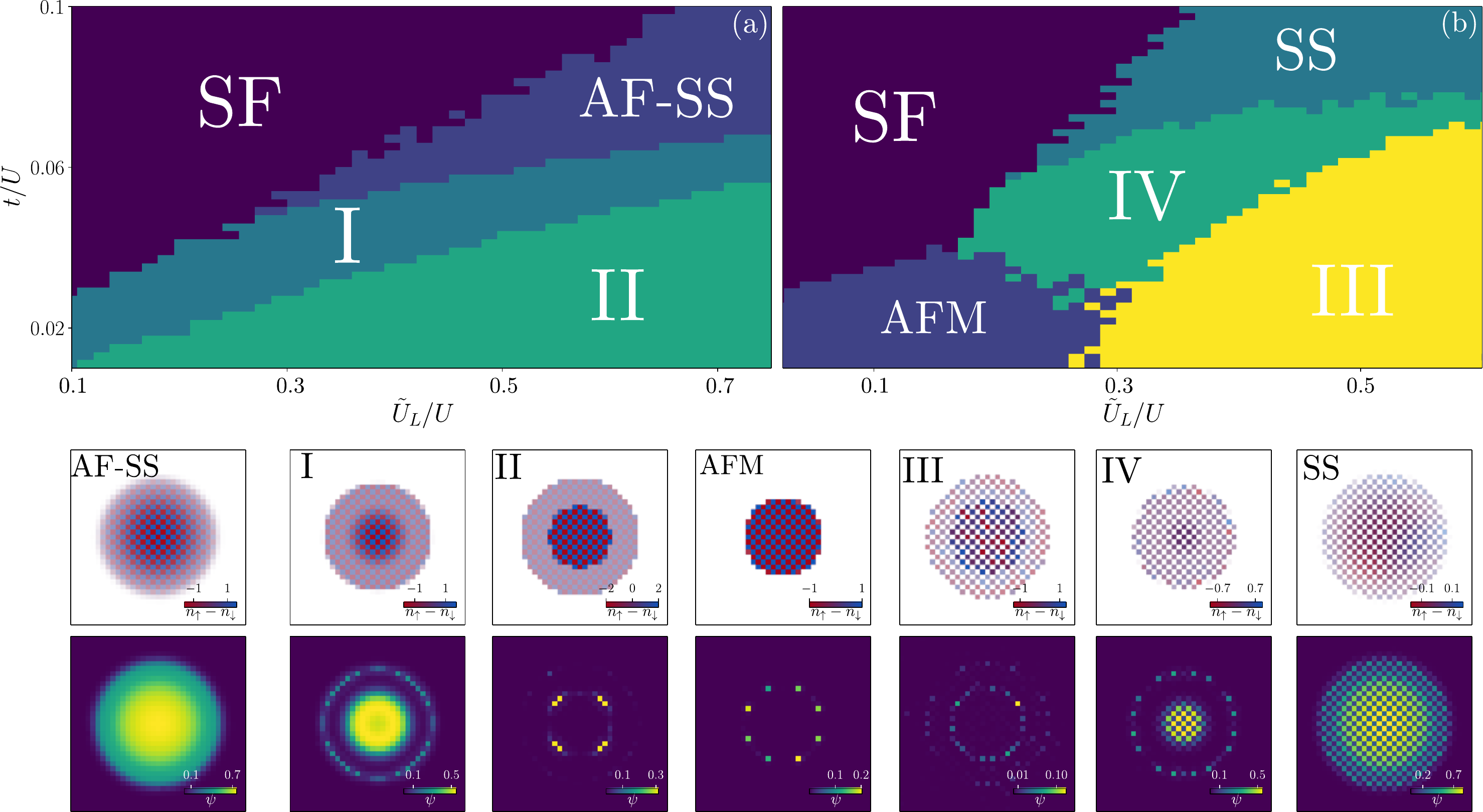}
			
			\caption{Behavior of the nonhomogeneous gas of spinor bosons coupled to a cavity in an optical lattice confined by an isotropic harmonic trap in 2D. The chemical potential and trapping strength in Eq.~(\ref{eq:harmonic}) are chosen as $\alpha = U / 100$ and $\mu = U$ in (a) and $\mu = U / 2$ in (b). Total zero magnetization is assumed with $P = 100 U  \times K$ and $M_{\textrm{tot}} = 0$.  Subplots (a) and (b) show the numerically obtained phase diagrams: (a) corresponds to predominantly vectorial cavity interactions, $\tilde{U}_s / \tilde{U}_v = 1/5$, while (b) corresponds to predominantly scalar interactions, $\tilde{U}_s / \tilde{U}_v = 5$. The phases are indicated by distinct colors and acronyms defined in Tab.~\ref{tab:trap_phases}. The bottom rows illustrate representative real-space configurations of the trapped phases on a $K = 35 \times 35$ lattice in terms of two observables: local magnetization $n_{\uparrow} - n_{\downarrow}$ (top panels) and the superfluid order parameter $\psi = (|\psi_{\uparrow}| + |\psi_{\downarrow}|)/2$ (bottom panels). The first three panels correspond to points in the parameter space of subplot (a): AF-SS: $\tilde{U}_L = 0.4 U,\ t = 0.078 U$; I: $\tilde{U}_L = 0.2 U,\ t = 0.039 U$;  II: $\tilde{U}_L = 0.48 U,\ t = 0.02 U$. The remaining panels correspond to subplot (b): AFM: $\tilde{U}_L = 0.25 U,\ t = 0.01U$;  III: $\tilde{U}_L = 0.5 U,\ t = 0.01 U$;  IV: $\tilde{U}_L = 0.27 U,\ t = 0.05 U$;  SS: $\tilde{U}_L = 0.5 U,\ t = 0.1 U$. AFM label corresponds         to the AFM phase with $\rho = 1$. The fluctuations of $\hat{n}_{\uparrow} - \hat{n}_{\downarrow}$ in subplots III, IV, and SS are due to the fact that the minimization landscape is flat in the spin direction for the considered zero-magnetization problem.}
			\label{fig:harmonic}
		\end{figure*}

		\section{Harmonic trapping}
		\label{sec:harm}

		In a realistic experiment,  one typically starts with an atomic cloud in which the ratio of spin-up to spin-down atoms is fixed, as spin-changing processes in such setups are usually negligible~\cite{Landini18}. As a consequence, phase transitions between states with different total magnetization are not directly accessible under these conditions. To account for this, we introduced Eq.~(\ref{eq:energy_func}) and analyzed the system with a soft constraint that fixes the total magnetization of the system in the previous sections. Additionally, we extend this setup by incorporating a harmonic trapping potential. The modified Hamiltonian in Eq.~(\ref{eq:energy_func}) reads
		\begin{equation}
			\hat{H}_{GC} = \hat{H} - \sum_{i=1}^{K}\sum_{\sigma \in \{ \uparrow, \downarrow \}} \left( \mu - E_i \right) \hat{n}_{i, \sigma}
			\label{eq:harmonic}
		\end{equation}
		where $E_i = \alpha[(i_x- \left\lceil L/2 \right\rceil)^2+(i_z-\left\lceil L/2 \right\rceil)^2]$, $\alpha$ denotes the strength of the harmonic trapping centered in the middle of the lattice, and $\left\lceil x \right\rceil$ denotes the ceiling of the number $x$. For simplicity and to reduce the number of parameters, we assume an isotropic harmonic potential. Still, because the trap term breaks the translational invariance of the Hamiltonian, we can no longer employ a two-site unit-cell Gutzwiller ansatz. Instead, we must consider a nonuniform Gutzwiller wavefunction, in which independent sets of coefficients are determined for each lattice site:
		\begin{equation}
			\left|\Psi \right> = \prod_{i=1}^{K}  \left( \sum_{n=0}^{n_{\textrm{max}}} \sum_{m=0}^{m_{\textrm{max}}} g_i(n, m) \left|n,m\right>_i \right).
			\label{eq:trap_gutz}
		\end{equation}
		As already pointed out for the single-component model in ~\cite{Li13} and \cite{Sundar16}, the extensive nature of the cavity-mediated interactions prevents the infinite-system solution from serving as a reliable approximation within a local density approximation. The system size is no longer an appropriate control parameter in the Hamiltonian~(\ref{eq:ham}), while the total particle number becomes the relevant quantity. We therefore rescale the cavity interaction strengths as
		\begin{equation}
			\tilde{U}_s = N U_{s} / K, \quad \tilde{U}_v = N U_{v} / K, \quad \tilde{U}_L = N U_L / K,
		\end{equation}
		where $N = \sum_{i, m , n}  g_{i}(n, m)^2 (m + n) $ is a total number of particles. The value of $N$ is determined self-consistently, i.e., it is not treated as an external parameter but enters directly into the minimization cost function (\ref{eq:energy_func})  - see Appendix~\ref{app:nonuniform_gutz}.
		
		It is worth noting that the symmetry-breaking structure of the CDW and AFM ground states prevents us from exploiting most of the residual geometric symmetries that may persist for an optical lattice in a harmonic trap~\cite{Zakrzewski05}. Therefore, we cannot significantly reduce the computational complexity using symmetry arguments, and we restrict ourselves to system sizes of $K = 35 \times 35$, truncating the local Hilbert space to a maximum occupation of $n_{\textrm{max}} = m_{\textrm{max}} = 6$ bosons per spin component.
		
		\begin{table}
			\centering
			\renewcommand{\arraystretch}{1.2} 
			\begin{tabular}{|l|l|}
				\hline
				Acronyms & Combination of phases in the trap \\
				\hline\hline
				I &  AF-SS  -- AFM$_{\rho=1}$  -- AF-SS \\
				II &  AF-SS  -- AFM$_{\rho=2}$  -- AF-SS  -- AFM$_{\rho=1}$ \\
				III & SS  -- CDW$_{\rho=1}$  -- SS  -- CDW$_{\rho=1/2}$ \\
				IV & SS  -- CDW$_{\rho=1/2}$  -- SS  \\
				
				\hline
			\end{tabular}
			\caption{Acronyms of the phases possible in the harmonic trap and the underlying combinations of the homogeneous phases - see Tab.~\ref{tab:phases}. In our convention, the phases are listed in order of increasing distance from the trap center, such that the rightmost term in the sequence represents the center region. underlying states are visually illustrated in Fig.~\ref{fig:harmonic}. 
			}
			
			\label{tab:trap_phases}
		\end{table}

		{In Fig.~\ref{fig:harmonic} we present the phase diagram obtained from the Hamiltonian~(\ref{eq:harmonic}) in the $(\tilde{U}_L / U, t / U)$ plane for the trapping strength $\alpha = U/100$ and zero total magnetization. The chemical potential is set to $\mu = U$ in Fig.~\ref{fig:harmonic}(a) and $\mu = U/2$ in Fig.~\ref{fig:harmonic}(b). As in the previous sections, Fig.~\ref{fig:harmonic}(a) corresponds to the case of dominant vectorial interactions, $\tilde{U}_s / \tilde{U}_v = 1/5$, while Fig.~\ref{fig:harmonic}(b) represents the case of dominant scalar interactions, $\tilde{U}_s / \tilde{U}_v = 5$. The bottom panels show representative color maps of the phases appearing in the diagram, highlighting the local magnetization, $n_{\uparrow} - n_{\downarrow}$, and the superfluid order parameter $\psi = (|\psi_{\uparrow}| + |\psi_{\downarrow}|)/2$ at each lattice site.} Additionally, all phases observed in the trapped system simulations are summarized in Tab.~\ref{tab:trap_phases}.  
		Note that (\ref{eq:harmonic}) employs the grand-canonical description, so the particle number fluctuates across the presented phase diagrams. However, for the one-component model with cavity, it was shown \cite{Sundar16} that such an approach reproduces the experimental data with good accuracy.

		The problem of identifying phases in the trap may be treated as a textbook pattern recognition \cite{Dawid25}. Each point in the ($\tilde{U}_L / U$, $t /U$) parameter space corresponds to a trapped system characterized by three observables on a lattice (depicted in the bottom rows of Fig.~\ref{fig:harmonic}): $n_{\uparrow}$, $n_{\downarrow}$, and $\psi$. By manually identifying and labeling a subset of points according to their phases, we construct a training set for a shallow convolutional neural network (CNN) \footnote{We employed the PyTorch Python library, creating a CNN with four learnable layers: two convolutional layers for spatial feature extraction and two fully connected layers for final classification. To prevent overfitting, the model was trained using a standard cross-entropy loss function on a dataset artificially expanded via 90$^{\circ}$ spatial rotations and the injection of Gaussian noise.} This CNN then classifies the remaining points, and the resulting phase regions are shown as areas with different colors in Fig.~\ref{fig:harmonic}(a) and (b). We stress that the CNN was utilized strictly for the labeling of existing configurations, not for the creation of new data points or interpolation between them.

		For the chosen trap parameters and dominant vectorial interactions, we identify four different regimes, shown in Fig.~\ref{fig:harmonic}(a). For small cavity-induced interactions and sufficiently large tunneling, the system is in the standard superfluid phase (SF), characterized by the absence of spatial or spin ordering and a finite superfluid order parameter. When $\tilde{U}_L$ is slightly increased, the system enters the trapped AF-SS phase, where, in addition to superfluid behavior, antiferromagnetic order (vectorial imbalance) emerges. Particles arrange themselves in a checkerboard pattern, with even and odd sites predominantly occupied by bosons with opposite spins. This can be seen in Fig.~\ref{fig:harmonic} (AF-SS), where the local magnetization has opposite signs on even and odd sites. Next, we identify phase I, corresponding to a wedding-cake structure in which the central region forms an  AF-SS, surrounded by the AFM phase with density $\rho = 1$. For sufficiently strong cavity-induced interactions, the system enters the final phase II, consisting of two AFM shells: a central region with density $\rho = 2$ and an outer region with $\rho = 1$.
		
		For the opposite configuration with stronger scalar interactions, we choose a smaller value of the chemical potential, $\mu = U/2$, to target the relevant densities over the full range of considered $\tilde{U}_L / U$. For small $\tilde{U}_L / U$, the system first exhibits an AFM phase (with density $\rho = 1$), in agreement with the qualitative expectations drawn from the homogeneous results. In contrast to the previous case, we do not observe the formation of an AF-SS phase between the AFM and SF phases as tunneling increases; instead, the transition to the SF phase is direct.
		
		As the strength of the cavity-mediated interactions increases, the system enters trapped phase III, which is characterized by shells of density-modulated checkerboard phases with particle densities $\rho = 0.5$ and $\rho = 1$. In other words, one observes a wedding cake of CDWs of different densities. We note that the fluctuations of $\hat{n}_{\uparrow} - \hat{n}_{\downarrow}$ seen in Fig.~\ref{fig:harmonic}(III) are due to the fact that the considered zero-magnetization problem is fully symmetric with respect to spin for $U_{12} = U$. As a consequence, the landscape is completely flat in the spin direction, and the optimizer chooses an arbitrary value. Therefore, as discussed in Section~\ref{sec:atom} and Section~\ref{sec:mean} for the homogeneous case, only density order has a physical meaning in this region. Increasing the tunneling within phase III leads to the development of superfluid order, forming a partial SS phase at the center of the trap; we denote this regime as phase IV. Finally, for larger tunneling strengths, the full SS phase emerges.

		\section{Conclusions}
		\label{sec:con}

		As experimental trapping of ultracold spinor bosonic gases in high-finesse optical cavities continues to advance, there is a growing need for theoretical studies of the corresponding extended Bose–Hubbard models with two-component bosons. Here, we investigate the simplest case of cavity-induced interactions that arise when the nodes of the optical lattice coincide with the antinodes of the cavity field, as originally introduced in \cite{Carl23}. We analyze the system within mean-field theory in the grand-canonical ensemble, considering the full unconstrained problem and introducing a zero-magnetization penalty. In the atomic limit, the unconstrained homogeneous system hosts two insulating phases: AFM and FDW. {When tunneling is introduced, these phases are surrounded by a superfluid and two supersolid phases, distinguished by different patterns of spin and density imbalances between odd and even sites. Enforcing the total magnetization to be zero leads to the transformation of the FDW phases into CDW states, yielding a density wave characterized by the absence of any magnetic ordering on the occupied lattice sites. Finally, we include a harmonic trapping potential in our simulations, which allows us to obtain the full phase diagram directly relevant for experiments, revealing a rich variety of trapped shell structures composed of phases observed for the homogeneous case.

			\acknowledgments
			We thank Tobias Donner and Tilman Esslinger for the suggestion of this study and the subsequent discussions.
			This research was carried out and financed within the framework of the second Swiss Contribution MAPS
			(Grant No. 230870). We gratefully acknowledge the Polish high-performance computing infrastructure PLGrid (HPC Center: ACK Cyfronet AGH) for providing computer facilities and support within the computational grant no. PLG/2025/018400.

			\bibliography{refs}
			
			\appendix
			\onecolumngrid
			
			\section{Atomic limit details}
			\label{app:atomic_limit}
			{
				In the atomic limit, the expectation value of the Hamiltonian (\ref{eq:grand}) in any product state has the following form
				
				\begin{equation}
					\begin{aligned}
						\left< \hat{H}_{GC}(t = 0) \right> 
						&=  \frac{U}{2} \sum_{i=1}^{K} \sum_{\sigma \in \{\uparrow, \downarrow \}} \left( \left<\hat{n}_{i, \sigma}^2 \right> - \left<\hat{n}_{i, \sigma} \right> \right) + U_{12} \sum_{i=1}^{K} \left<\hat{n}_{i, \uparrow} \hat{n}_{i, \downarrow}\right>  - \mu \sum_{i=1}^{K} \sum_{\sigma \in \{\uparrow, \downarrow \}} \left< \hat{n}_{i, \sigma}\right> \\
						&\quad - \frac{U_s}{K} \left\{ \sum_{i=1}^{K} \left( \left<(\hat{n}_{i, \uparrow} +\hat{n}_{i, \downarrow})^2\right>  - \left<\hat{n}_{i, \uparrow} +\hat{n}_{i, \downarrow}\right>^2 \right)  + \left(\sum_{i=1}^K (-1)^i \left<\hat{n}_{i, \uparrow} +\hat{n}_{i, \downarrow} \right> \right)^2 \right\} \\
						&\quad  - \frac{U_v}{K} \left\{ \sum_{i=1}^{K} \left( \left<(\hat{n}_{i, \uparrow} -\hat{n}_{i, \downarrow})^2\right>  - \left<\hat{n}_{i, \uparrow} -\hat{n}_{i, \downarrow}\right>^2 \right)  + \left(\sum_{i=1}^K (-1)^i \left<\hat{n}_{i, \uparrow} -\hat{n}_{i, \downarrow} \right> \right)^2 \right\}
					\end{aligned}
				\end{equation}
				To find the ground states in the homogeneous unconstrained case, it is sufficient to assume that the solutions are Fock states spanning the infinite lattice $K \rightarrow \infty$ with a two-site unit cell structure. In other words, we are solving for the states (\ref{eq:uniform_gutz}) with $\ket{\phi_e}_{2i} = \ket{n_{e, \uparrow}, n_{e, \downarrow}}$,  $ \ket{\phi_o}_{2i-1} = \ket{n_{o, \uparrow}, n_{o, \downarrow}}$, where $n_{i,m}$ are the four corresponding occupation numbers. The ground state is therefore obtained by minimizing the following energy density over these integers:
				
			}
			
			\begin{equation}
				\begin{aligned}
					\mathcal{E}_0
					&= \frac{2}{K}\braket{\hat{H}_{GC}(t=0)}  = \frac{U}{2} \sum_{\sigma  \in\{\uparrow, \downarrow\}} \left(n_{e, \sigma} (n_{e, \sigma} - 1)+ n_{o, \sigma} (n_{o, \sigma} - 1)\right) + U_{12}\left(n_{e, \uparrow}n_{e, \downarrow} + n_{o, \uparrow}n_{o, \downarrow} \right) \\
					&\quad - \frac{U_s}{2}\left(n_{e, \uparrow} + n_{e, \downarrow} - n_{o, \uparrow} - n_{o, \downarrow} \right)^2 - \frac{U_v}{2}\left(n_{e, \uparrow} - n_{e, \downarrow} - n_{o, \uparrow} + n_{o, \downarrow} \right)^2 \\
					&\quad  - \mu\sum_{\sigma  \in\{\uparrow, \downarrow\}} \left( n_{e, \sigma} + n_{o, \sigma} \right),
				\end{aligned}
			\end{equation}
			
			For analytical convenience, it is useful to rewrite this expression in terms of new variables. We introduce the scalar (density) imbalance and vectorial (spin) imbalance
			
			$$ \theta_s = n_{e, \uparrow} + n_{e, \downarrow} -  n_{o, \uparrow} - n_{o, \downarrow}, \quad \theta_v = n_{e, \uparrow} - n_{e, \downarrow} -  n_{o, \uparrow} + n_{o, \downarrow},$$
			as well as the spin-resolved densities
			
			$$ \rho_{\uparrow} = \frac{n_{e, \uparrow} + n_{o, \uparrow}}{2}, \quad \rho_{\downarrow} = \frac{n_{e, \downarrow}   + n_{o, \downarrow}}{2}, $$
			which sum to the total density $\rho =  \rho_{\uparrow} + \rho_{\downarrow}$, (\ref{eq:density}). 
			
			In these variables, the energy density takes the form
			
			\begin{equation}
				\mathcal{E} = \frac{\mathcal{E}_0}{2} = \frac{U}{2} \left[ \rho_{\uparrow} (\rho_{\uparrow} - 1) + \rho_{\downarrow} (\rho_{\downarrow} - 1) \right] + U_{12} \rho_{\uparrow} \rho_{\downarrow}  - \mu ( \rho_{\downarrow} + \rho_{\uparrow})  \\+ \left(\frac{U + U_{12}}{4} - U_s \right) \frac{\theta_s^2}{4}  + \left(\frac{U - U_{12}}{4} - U_v \right) \frac{\theta_v^2}{4}.
				\label{eq:app:atomic_limit}
			\end{equation}
			The ground state is obtained by minimizing this expression with respect to the integer occupations $n_{i,m}$. Assuming $U = U_{12}$, one recovers the formula given in the main text, (\ref{eq:atomic_limit}).

			\section{Homogeneous Gutzwiller ansatz}
			\label{app:hom_gutz}
			
			The energy density of the model in the homogeneous case, assuming the Gutzwiller ansatz~(\ref{eq:uniform_gutz}), takes the following form:

			\begin{equation}
				\begin{aligned}
					\varepsilon(g_e, g_o)
					&=  -2  z  t \left(\sum_{n,m} \sqrt{n} g_e(n ,m) g_e(n-1, m) \right)  \left(\sum_{n,m} \sqrt{n} g_o (n ,m) g_o(n-1, m) \right)  \\
					&\quad
					-2  z  t \left( \sum_{n,m} \sqrt{m} g_e (n ,m) g_e(n, m-1) \right) \left(\sum_{n,m} \sqrt{m} g_o (n ,m) g_o(n, m-1) \right)\\
					&\quad
					+ \frac{U}{2}\sum_{i \in\{e, o\}} \sum_{n,m} g_i(n, m)^2 \left(n(n-1) + m(m-1) \right) + U_{12} \sum_{i \in\{e, o\}} \sum_{n,m} g_i(n, m)^2 n ~m  \\ 
					&\quad
					- \frac{U_s}{2} \left( \sum_{n,m} (g_e(n,m)^2 - g_o(k,l)^2) (n + m) \right)^2 - \frac{U_v}{2}  \left(\sum_{n,m} (g_e(n,m)^2 - g_o(k,l)^2) (n - m) \right)^2\\
					&\quad
					- \mu \sum_{i \in\{e, o\}} \sum_{n,m} g_i(n, m)^2 \left(n + m \right) + h \sum_{i \in\{e, o\}} \sum_{n,m} g_i(n, m)^2 \left(n - m \right)\\
					&\quad
					+ \frac{P}{2} \left(\sum_{i \in\{e, o\}} \sum_{n,m} g_i(n, m)^2 \left(n - m \right) - 2 M_{\textrm{tot}} \right)^2
					\label{eq:gutz_func_hom}
				\end{aligned}
			\end{equation}
			{This expression is obtained in the limit of $K \rightarrow \infty$}, reducing the problem to an optimization on a two-site unit cell with two vectors of variational parameters, $g_e$ and $g_o$. We perform the minimization using the L-BFGS algorithm implemented in the SciPy numerical library~\cite{Virtanen20}, without explicit constraints, and compute the analytical gradients directly from the expression above, $\partial \varepsilon / \partial g_i$.
			
			To enforce the normalization of the Gutzwiller coefficients, $\|g_i\|^2 = 1$, we allow the optimizer to explore the full parameter space, while normalizing vectors in the cost function. Thus, the gradients must be projected along this normalized path.

			\section{Nonuniform Gutzwiller ansatz in the presence of the trapping potential}
			\label{app:nonuniform_gutz}
			
			As discussed in the main text, the nonhomogeneous case requires the full lattice Gutzwiller ansatz~(\ref{eq:trap_gutz}), which leads to the following energy functional,
			\begin{equation}
				\varepsilon (\vec{g}) = \sum_{i=1}^{K}  \varepsilon_i(\vec{g}) + \frac{P}{K} \left( \sum_{i=1}^{K}\sum_{n,m} g_i(n, m)^2 \left(n - m \right)  - M_{\textrm{tot}} \right)^2,
			\end{equation}
			where the local contribution to the energy at each site is given by
			\begin{equation}
				\begin{aligned}
					\varepsilon_i(\vec{g})
					&=  -2  t   \left(\sum_{n,m} \sqrt{n} g_i(n ,m) g_i(n-1, m) \right)   \sum_{j \in \left< i \right>}  \left(\sum_{n,m} \sqrt{n} g_j(n ,m) g_j(n-1, m) \right) \\
					&\quad
					-2  t   \left(\sum_{n,m} \sqrt{n} g_i(n ,m) g_i(n, m-1) \right)   \sum_{j \in \left< i \right>}  \left(\sum_{n,m} \sqrt{n} g_j(n ,m) g_j(n, m-1) \right)\\
					&\quad
					+ \frac{U}{2} \sum_{n,m} g_i(n, m)^2 \left(n(n-1) + m(m-1) \right) + U_{12} \sum_{n,m} g_i(n, m)^2 n ~m \\
					&\quad
					- \tilde{U}_s\frac{ \sum_{n,m}\left( g_i(n,m)^2 (n +m)^2 +\sum_{k,l} \sum_{j \neq i} (-1)^{|i| + |j|} g_i(n,m)^2 g_j(k,l)^2 (n +m)(k +l) \right)}{ \sum_{n, m} g_i(n, m)^2 \left(m + n \right) } \\
					&\quad
					-\tilde{U}_v \frac{\sum_{n,m} \left( g_i(n,m)^2 (n - m)^2 +\sum_{k,l} \sum_{j \neq i} (-1)^{|i| + |j|} g_i(n,m)^2 g_j(k,l)^2 (n - m)(k - l) \right) }{ \sum_{n, m} g_i(n, m)^2 \left(m + n \right)} \\
					&\quad
					+ \sum_{n, m} \left(E_i - \mu \right) g_i(n, m)^2 \left(n + m \right) 
					\label{eq:gutz_func_trap}
				\end{aligned}
			\end{equation}
			where $\vec{g}$ is constructed from the Gutzwiller coefficients corresponding to the sites of the entire lattice, and $\langle i \rangle$ denotes the set of nearest neighbors of site $i$. As discussed above, the total particle number enters the denominators of the cavity-induced interaction terms, reflecting the extensive character of these interactions.
			
			For the nonuniform case, we employ the same optimization routine as described in Appendix.~\ref{app:hom_gutz} for the homogeneous situation.
			
		\end{document}